\documentclass[12pt]{iopart}

\usepackage{epsfig}
\usepackage{amssymb}
\usepackage{setstack}
\usepackage{iopams}
\usepackage{subfigure}

\newcommand{\Bv}{\mathbf{B}}
\newcommand{\Bpol}{B_{\mathrm{p}}}
\newcommand{\bv}{\mathbf{b}}
\newcommand{\vv}{\mathbf{v}}
\newcommand{\jv}{\mathbf{j}}

\newcommand{\dott}{ \mbox{\boldmath\(\cdot\)}}
\newcommand{\cross}{ \mbox{\boldmath\(\times\)}}
\newcommand{\grad}{ \mbox{\boldmath\(\nabla\)}}
\newcommand{\divv}{ \mbox{\boldmath\(\nabla\cdot\)}}
\newcommand{\curl}{ \mbox{\boldmath\(\nabla\times\)}}
\begin{document}

\title[INCOMPRESSIBLE LIMIT FOR RESISTIVE PLASMAS]{
  A COMPARISON OF INCOMPRESSIBLE LIMITS FOR
  RESISTIVE PLASMAS.}

\author{B.F. McMillan, R.L. Dewar}
{Department of Theoretical Physics,
Research School of Physical Sciences and Engineering, The Australian
National University, Canberra 2600, Australia.}
\author{R.G. Storer }{School of Chemistry, Physics and Earth Sciences,
Flinders University, Adelaide 5001, Australia
}

\def\lsimeq{\;\raise.6ex\hbox{$<$}\mskip-13.5mu\lower.6ex\hbox{$\sim$}\;}

\begin{abstract}

\newcommand{\iotabar}{\iota \!\!\! - }

The constraint of incompressibility is often used
to simplify the magnetohydrodynamic (MHD) description of linearized
plasma dynamics because it does not affect the ideal MHD marginal
stability point.
In this paper two methods for introducing
incompressibility are compared in a cylindrical plasma model:
In the first method,
  the limit
$\gamma \rightarrow \infty$ is taken, where $\gamma$ is the ratio of
specific heats; in the second, an anisotropic mass tensor $\mathbf{\rho}$ is
used, with the component parallel to the magnetic field taken to vanish,
$\rho_{\parallel} \rightarrow 0$.
Use of resistive MHD reveals the nature of these two
limits because the Alfv\'en and slow magnetosonic continua of ideal MHD
are converted to point spectra and moved into the complex plane. Both limits
profoundly change the slow-magnetosonic spectrum, but
only the second limit faithfully reproduces the resistive Alfv\'en
spectrum and its wavemodes.
In ideal MHD, the slow magnetosonic continuum
degenerates to the Alfv\'en continuum in the first method,
while it is moved to infinity by the second.
The degeneracy in the first is broken by finite resistivity.
For numerical and semi-analytical study of these models,
we choose plasma equilibria which cast light on puzzling aspects of
results found in earlier literature.
\end{abstract}.

\maketitle

\section{Introduction}

We devote our attention to the ideal and resistive MHD models, which
despite their dramatic simplification of plasma behaviour,
are crucial to the design and operation of controlled fusion devices
and are at the core of many astrophysical plasma models.
Simplified models such as these have utility if they
can be used to make testable
predictions or if they yield insight into the internal processes of a
system. We test two variants of incompressible resistive MHD with these
criteria in mind.
The starting point for the analysis of most plasma models is an
understanding of the wavemodes that arise: this provides
information about the linear response and stability of the system and
provides a basis for much nonlinear analysis.
We focus on the linear behaviour of the plasma in this paper.
For the resistive MHD model, which includes dissipation, the wavemodes
are non-normal:
a full picture of linear plasma behaviour requires
an analysis of the transient behaviour of the system, as well as the
eigenvalue analysis which predicts asymptotic behaviours over long
time scales. These are closely related via pseudospectral
methods \cite{Pseudo}. In this paper we restrict attention to
eigenvalue analysis.

\noindent For many plasmas of physical interest it is true that the 
resistive term
is small: typically this is quantified by a large magnetic Reynolds number.
It might have been expected that for small enough resistivity,
the resistive MHD model could simply be treated as a perturbation of the
ideal MHD model. However, the change induced is actually a singular
perturbation, which introduces higher spatial derivatives.
One of the interesting effects of this property is that
eigenfrequencies in the ideal model are not necessarily approached
by the eigenfrequencies of any resistive modes, even for vanishingly
small resistivity.

\noindent Many papers have been published on the stable resistive MHD spectrum
and several of the early papers
(\cite{Davies}-
\nocite{DewarDavies}
\nocite{RyuGrimm}
\nocite{KernerLerbinger}
\cite{StorerAnalytic})
focused on cylindrical models.
These papers
have established certain generic features of the resistive spectrum.
The resistive spectrum is discrete, unlike the ideal MHD spectrum which
has continua: on some intervals, every frequency corresponds to a
generalised wavemode.
In the resistive spectrum, a large
number of fully complex eigenfrequencies can be found,
and in general these lie along loci, or lines, on the complex
plane. Generally as the resistivity is decreased to zero these lines
become densely populated with eigenvalues.

\noindent Resistive MHD is a simple closure of
the full kinetic equations, and as a result the plasma
dynamics parallel to the magnetic field lines are often quite
poorly represented \cite{Freidberg}.
For Alv\'enic modes, which do not strongly
compress the plasma, these parallel dynamics are generally
unimportant. However, for
the slow and fast magnetoacoustic waves, the parallel dynamics
and the effects of compressibility are important; these waves are not
necessarily well modelled by resistive MHD.

\noindent It is possible to find the compressible
resistive MHD spectrum numerically (as in \cite{KernerLerbinger})
and ignore the slow and fast magnetoacoustic waves that
are present. On the other hand, there are
approaches which promise to isolate the
Alfv\'enic portion of the spectrum and simplify the analysis.
We present two of these incompressible
approximations, in which the predicted motions of the plasma
satisfy $\nabla \dott \vv = 0$ (at least approximately).
One approach is to artificially set the ratio
of specific heats $\gamma$ to infinity
(as in \cite{RyuGrimm} and \cite{StorerAnalytic}). In the other,
an anisotropic mass tensor $\mathbf{\rho}$ is
used, with the component parallel to the magnetic field taken to vanish,
$\rho_{\parallel} /\rho_{\perp} \rightarrow 0$. With this density tensor,
ideal eigenmodes are incompressible,
but to ensure exact incompressibility for resistive eigenmodes
$\gamma$ must again be set to $\infty$. We can view these models as
the extreme cases of a generalised resistive MHD model with
two parameters, $\gamma$ and
$\rho_{\parallel} / \rho_{\perp}$.
The two extreme cases are not equivalent, and the resulting spectra are
qualitatively different.
We investigate these two methods, and compare
them with the compressible resistive MHD model. We
specialise to equilibria with zero background flow.
Note that $\gamma \rightarrow \infty$ may be physically appropriate for
particular conductive fluids and plasmas with $\beta >> 1$.

\noindent First, we examine the plane waves of the homogeneous 
incompressible MHD
model. Then we evaluate spectra in a simple cylindrical equilibrium
for varying values of
$\gamma$, and with and without an artificial anisotropic density.
This illustrates the transition between the compressible and
incompressible cases. We then discuss the spectra of
more general plasma configurations.
A WKB analysis of a generic incompressible model
is then undertaken in order to understand the
features of these spectra and to verify the numerics.
We begin by
solving the dispersion relation. Then the singular
features of this function are explored by reducing it to
a simpler form. To complete the groundwork for semi-analytic
calculations, the behaviour of the wave equation near these singular
points is examined. Finally, we use our WKB analysis to find the spectrum of an
example case.

\section{Wavemodes in incompressible MHD limits}

The first step in the analysis of these incompressible limits is
a determination of the wavemodes in a simple homogeneous plasma.
To this end we follow \cite{Freidberg} and derive wave frequencies.
We begin by considering a wave with wavevector at some
angle to the magnetic field $\mathbf{B} = B_0 \hat{z} $, so
$\mathbf{k} = k_{\parallel} \hat{z} +k_{\perp} \hat{x}$,
travelling in a plasma with sound speed $V_s = (\gamma p_0 / \rho_0)^{1/2}$
and Alfv\'{e}n speed $V_a = (B_0^2 / \mu_0 \rho_0)^{1/2}$.
We recover the Alfv\'{e}n spectrum:
\begin{equation}
\omega_A^2 = k_{\parallel}^2 {V_a}^2,
\end{equation}
and also two other solutions to the plasma equations:
\begin{equation}
\omega_{\pm}^2 = \frac{1}{2} k^2 ({V_a}^2+{V_s}^2
     +\frac{\rho_{\parallel}-\rho_{\perp}}{\rho_{\parallel} k^2} 
k_{\parallel}^2 V_s^2)
       \left(1 \pm (1-\alpha^2)^{\frac{1}{2}} \right),
\end{equation}
where
\begin{equation}
\alpha^2 = \frac{4 \rho_{\perp} k_{\parallel}^2 {V_a}^2 {V_s}^2}
          {k^2 \rho_{\parallel} ({V_a}^2+{V_s}^2
         +\frac{\rho_{\parallel}-\rho_{\perp}}{\rho_{\parallel} k^2 } 
k_{\parallel}^2 V_s^2)^2}.
\end{equation}
In low-$\beta$ compressible plasmas, $\omega_{+}$ corresponds
to the fast magnetoacoustic wave, and
$\omega_{-}$ to the slow magnetoacoustic wave.

\noindent In the limit $\gamma \rightarrow \infty$ (with 
$\rho_{\parallel}/\rho_{\perp} =1$)
  we find
$\omega_{+}^2 \rightarrow \infty $ and
$\omega_{-}^2 \rightarrow k_{\parallel}^2 {V_a}^2$, so that the slow-mode
is now degenerate with the Alfv\'{e}n mode.
In more general plasma configurations,
the slow and the Alfv\'en wavemodes
still occur at very similar frequencies,
and therefore can be strongly mixed. We show this does occur,
so that generic spectra determined
are composed of an unphysical combination
of these types of wavemodes.
In the limit $\rho_{\parallel}/\rho_{\perp} \rightarrow 0$
we again have $\omega_{+} \rightarrow \infty $, but
$\omega_{-}^2 \rightarrow k^2 V_a^2 +k_{\perp}^2 V_s^2$, which
is slightly larger than the fast magnetoacoustic frequency. In this
case we have suppressed the slow magnetoacoustic waves.

\noindent If we set
$\rho_\parallel / \rho_\perp \rightarrow 0 $,
we can show from the linearised equations
that for general resistive MHD wavemodes
$\rho_\parallel \rightarrow 0 $ implies:
\begin{equation}
  \Bv_0 \dott \grad (\divv \vv ) =
  - \frac{ \grad P_0 \dott \curl (\eta \jv ) }{ (\gamma P_0) } ,
\end{equation}
where
$B_0$ and $P_0$ are the equilibrium field and pressure,
$\eta$ is the resistivity and $\jv$ and $\vv$ are the
perturbed current and velocity.
So for the ideal case ($\eta=0$) we have
that $\divv \vv$ is a constant on all irrational surfaces,
and, by continuity, for finite
toroidal or poloidal mode number, we must have $\divv \vv =0$.
In the resistive case, we have small $\eta$, but possibly
large $d/dr$ so that resistive modes are not strictly incompressible.
However, if we also require $\gamma \rightarrow \infty$ then the
resistive modes are strictly incompressible.

\section{Numerical results of varying incompressibility}
\label{sec:Numerical}

In order to show the effect of incompressibility on the resistive
MHD spectrum,
we solved the compressible,
resistive MHD equations numerically. 
We use a code based on the description in
\cite{Gruber}.

\begin{figure}
\centering
\begin{tabular}{@{}l@{}l@{}l@{}}
\subfigure[$\gamma= \frac{5}{3} $]{
  \epsfig{figure=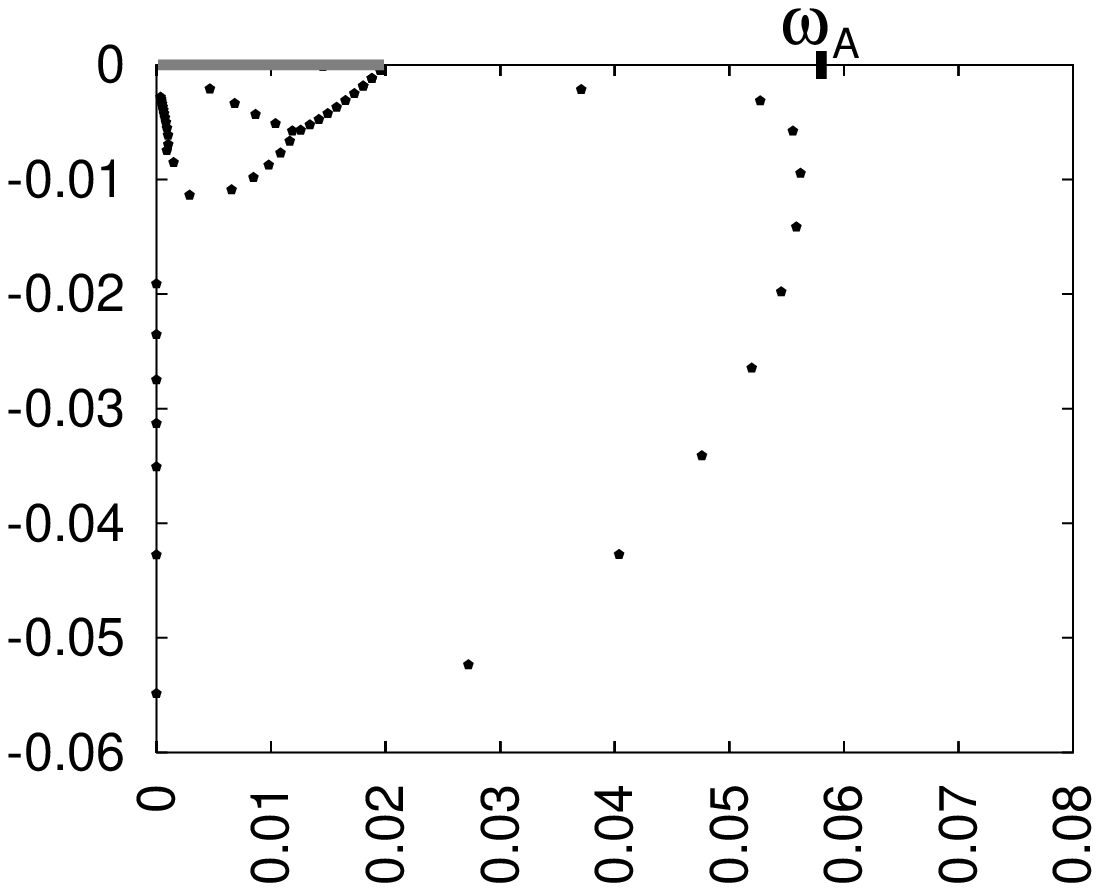,width=5cm}
  \label{figure:CompressTest2_gamma1}
                                           }   &
\subfigure[$\gamma=10 $]{
  \epsfig{figure=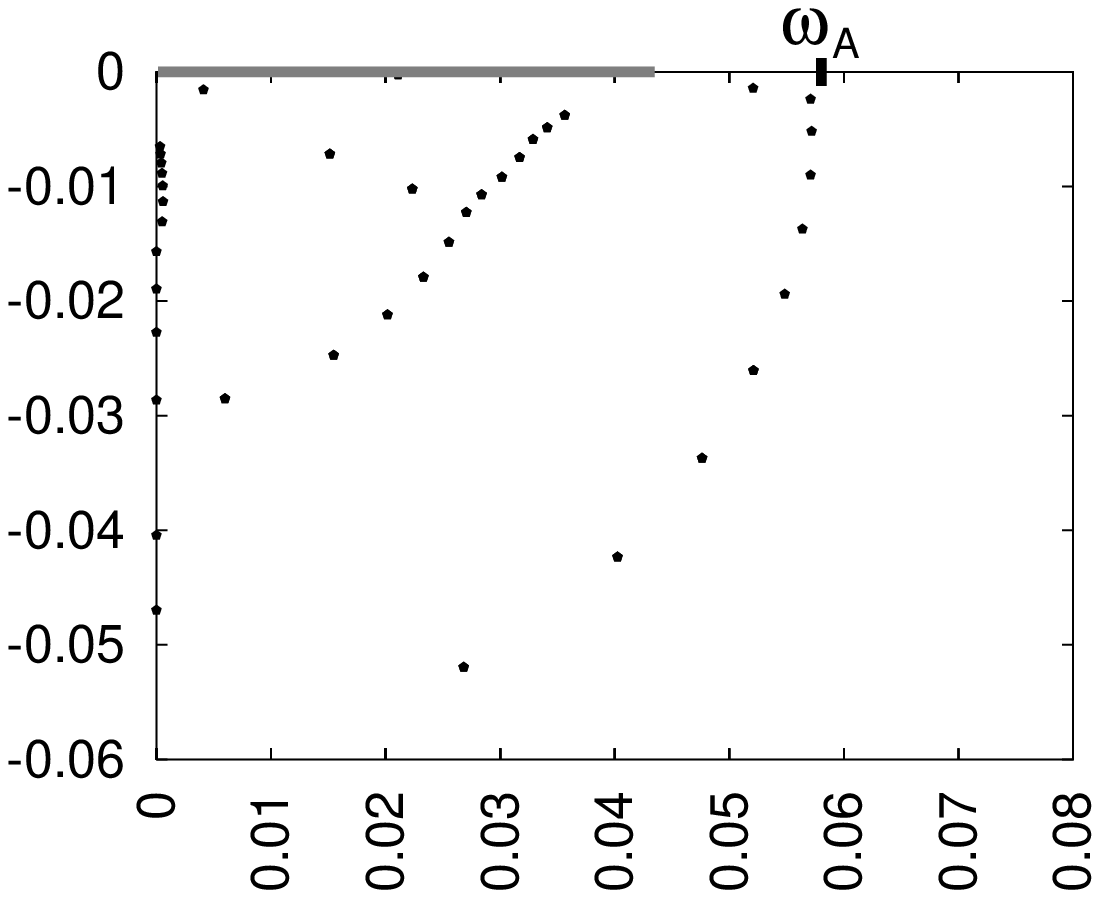,width=5cm}
  \label{figure:CompressTest2_gamma10}
}  &
\subfigure[$\gamma=20 $]{ \epsfig{figure=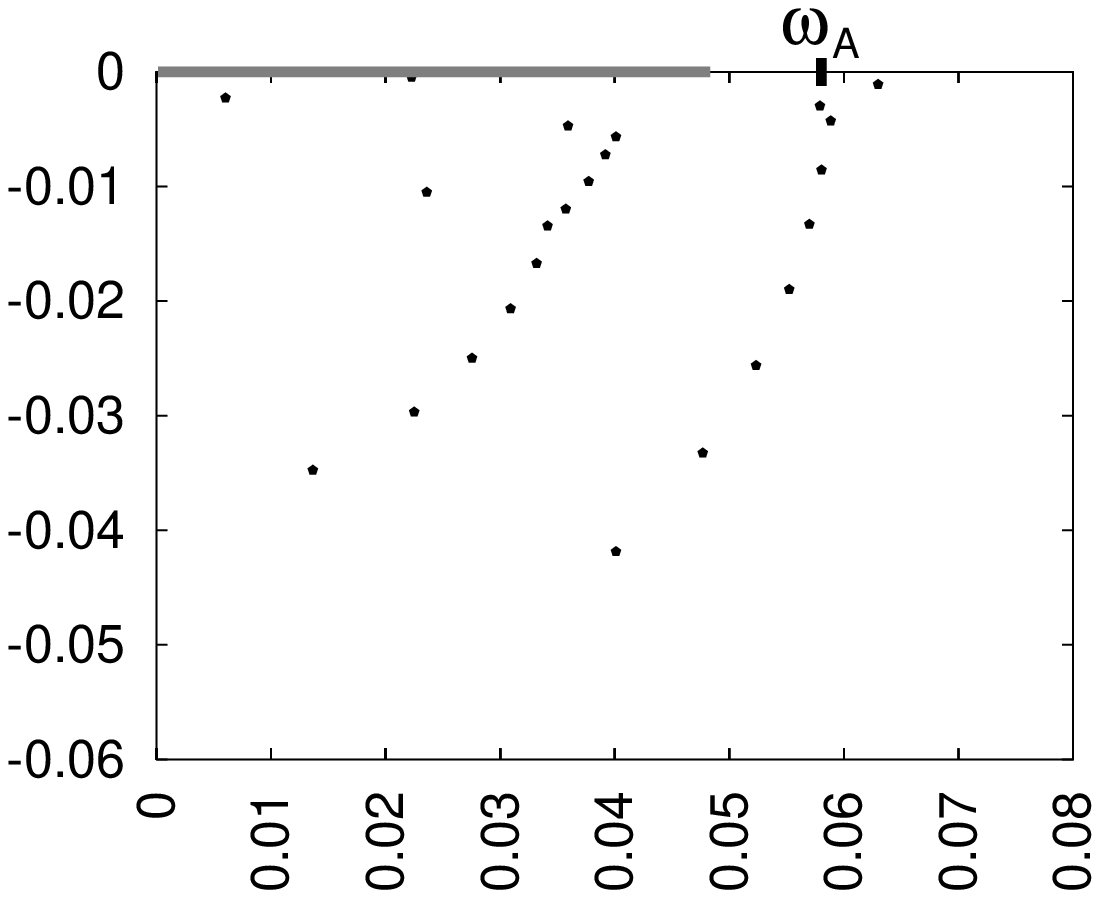,width=5cm} }  \\
\subfigure[$\gamma=40 $]{ \epsfig{figure=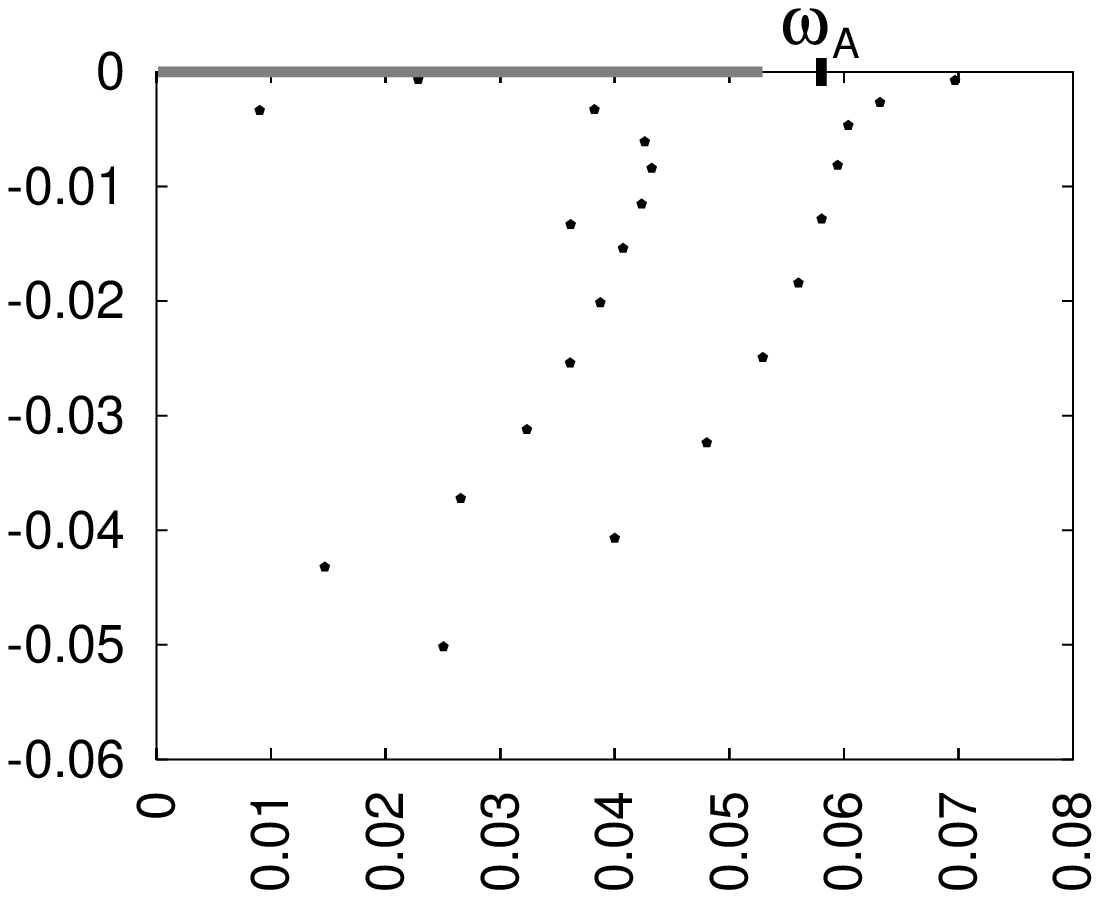,width=5cm} }  &
\subfigure[$\gamma=1000 $]{
   \epsfig{figure=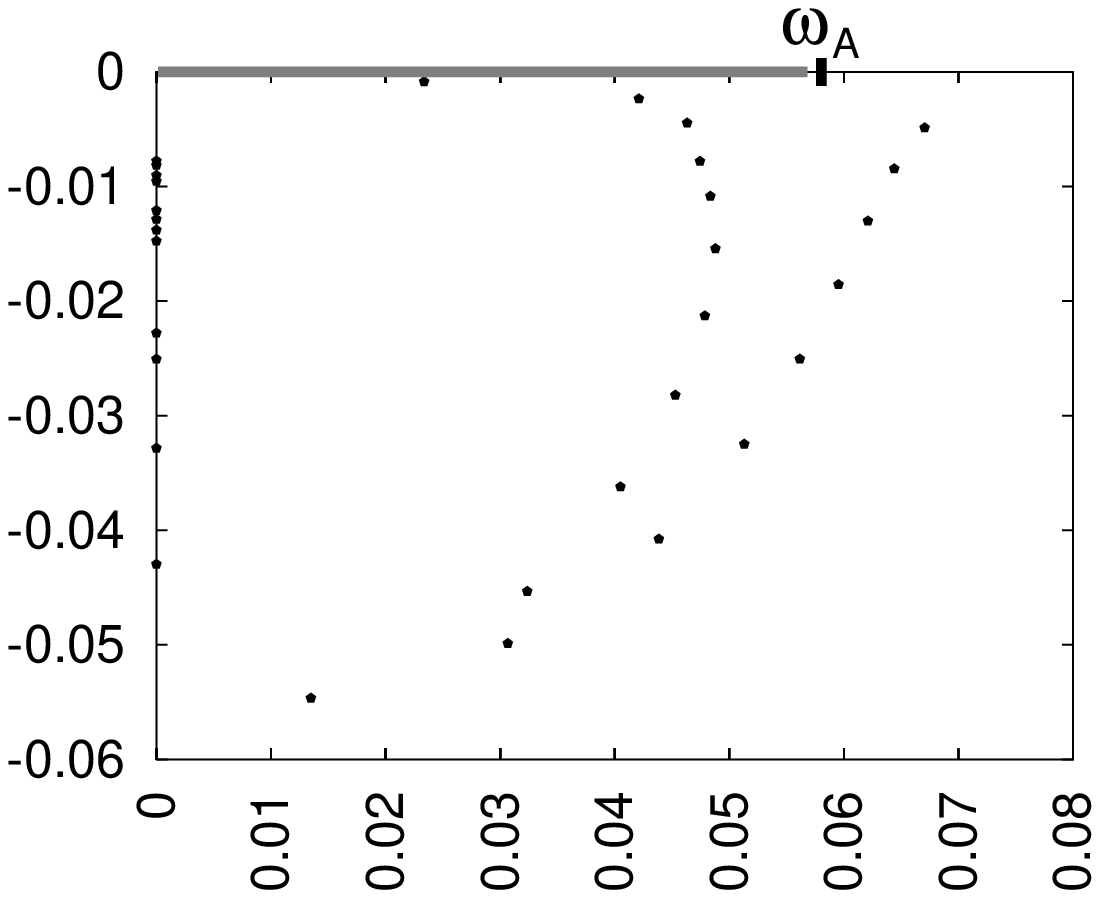,width=5cm}
   \label{figure:CompressTest2_gamma1000}
 } &
\subfigure[$\gamma=1000, \rho_{\parallel} \rightarrow 0$]{
   \epsfig{figure=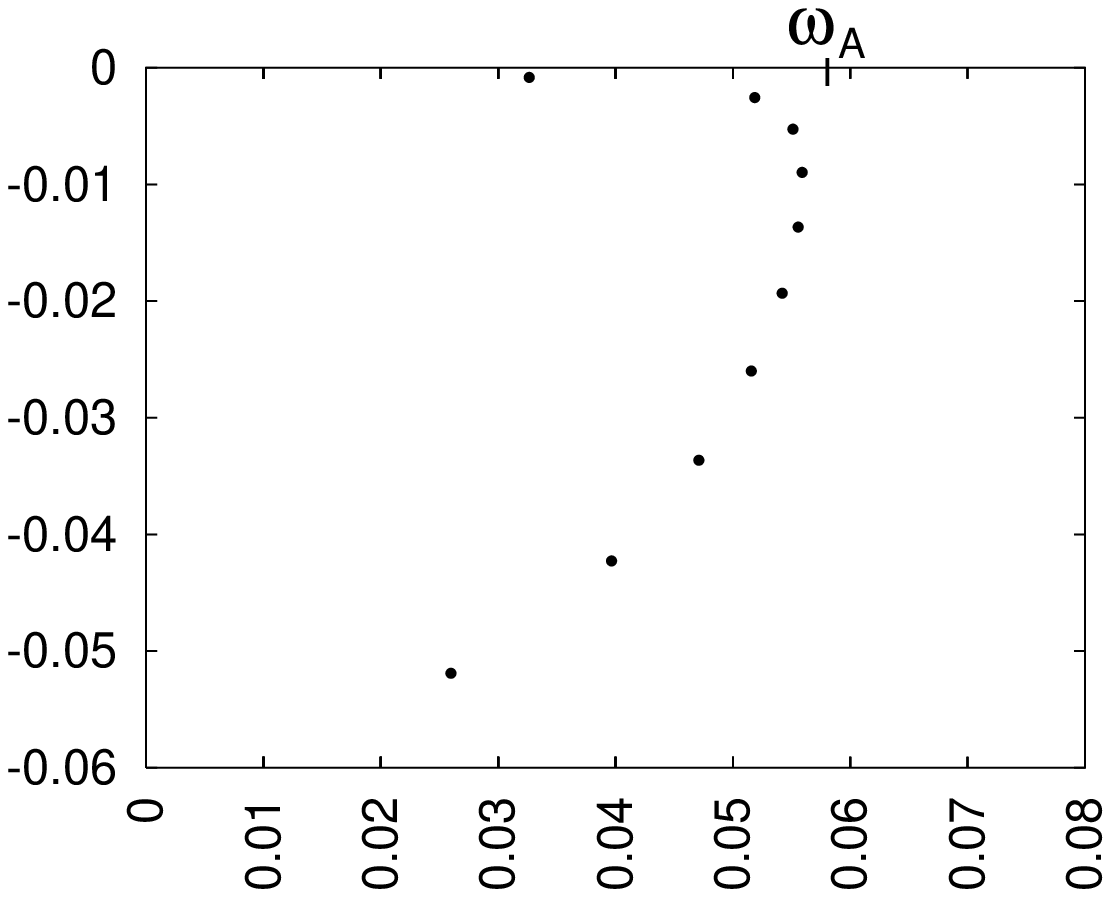,width=5cm}
   \label{figure:CompressTest2_rhop}
  }

\end{tabular}

\caption{The compressible resistive spectrum of a constant current
model ($\beta = 4\%$) for various values of $ \gamma $.
The ideal slow-mode continuum is represented by a
grey line on the real axis.
}
\label{figure:CompressTest2}
\end{figure}

We examine a cylindrical, zero-shear model case, as described in 
\cite{StorerAnalytic} , with $\beta \approx 4\%$.
The incompressibility is explored
by varying $\gamma$ in the range $1$ -- $1000$.
The incompressible limits correspond to
$\gamma \rightarrow \infty $, but in this case $\gamma \simeq 1000$
is high enough to demonstrate the limit. We define the magnetic 
Reynolds number $ S = \tau_R / \tau_A $
where $\tau_A $ and  $\tau_R $ are the Alfv\'{e}n and resistive
timescales. For a cylinder of radius $ r_p $ we have
$ \tau_A = r_p (\mu_0 \rho)^{1/2} / B_{z} $, and
$\tau_R=r_p^2 \mu_0 / \eta_0 $. The magnetic field perturbations are 
of the form $\bv = \exp \left(
i m \theta - i \kappa z / r_p - i \omega
t \right) \bv\left(r \right) $, with  $\kappa= n r_p / R$, by 
analogy with the toroidal case $R$ can
be interpreted in the sense that $ 2 \pi R $ is the length of the plasma
column and $n$ is the `toroidal' mode number. We have $\beta \approx 4\% $,
which allows the slow-mode spectrum to be shown on the same scale
as the Alfv\'{e}n spectrum in the compressible case.
The resulting spectra are shown in figure \ref{figure:CompressTest2}. 
For this case $m=1$, $\kappa=0.35$, $nq=1.2$ and $S=1\times10^4$.
Note that figure \ref{figure:CompressTest2_gamma1000} corresponds
to the limit $\gamma \rightarrow \infty $ and
figure \ref{figure:CompressTest2_rhop} corresponds to the limit
$\rho_{\parallel}/\rho_{\perp} \rightarrow 0$.

\noindent In these cases the ideal Alfv\'en continuum degenerates to 
a point, at
${\omega}_{A} = 0.057 $, but the ideal slow continuum is finite in extent
because of pressure and field strength variation across the plasma.
The slow continuum extends to the origin because the pressure is taken
to be zero at the plasma boundary.
Note the fork structure seen for the slow modes near the origin of
figure \ref{figure:CompressTest2_gamma1}. This fork structure is lost
as $\gamma$ is increased
[figures \ref{figure:CompressTest2_gamma10}--
   (e)].
Finally, as $\gamma \rightarrow \infty$, most of the mode frequencies
are in the
vicinity of a semicircle of radius $\omega_A$ on the complex plane.
 From the figure, we see that
there are many more modes near
$\omega_A$ in the $\gamma \rightarrow \infty$ model, than
in the more physical compressible model.
It has been shown in \cite{StorerAnalytic} that for this incompressible
case
wavemodes are eigenfunctions of helicity and none of the modes correspond
directly to physical compressible wavemodes.
In figure \ref{figure:CompressTest2_gamma1000}, the two loci of eigenvalues
correspond to wavemodes of opposite helicity.

\noindent For the
$\rho_{\parallel}/\rho_{\perp} \rightarrow 0$
model, we find a spectrum
[figure \ref{figure:CompressTest2_rhop}] very
similar to the compressible spectrum in figure
\ref{figure:CompressTest2_gamma1}, but with the notable absence
of the slow-mode fork. The position of individual Alfv\'enic
eigenvalues is in fact well preserved in this model.
The only noticable deviation is the eigenmode near the real axis,
at
$ Re( \omega ) \simeq 0.035 $, which has a frequency shift of
magnitude 
$ \simeq 0.004 $ as a result of setting $\rho_{\parallel}/\rho_{\perp}
\rightarrow 0$.
Since this Alfv\'en eigenmode is
fairly close in frequency to the slow modes, it is not surprising
that it is the one most strongly modified by an assumption of
incompressibility.

\section{Generic spectra in resistive MHD }

\begin{figure}[htb]
\centerline{\epsfxsize=8cm \epsfbox{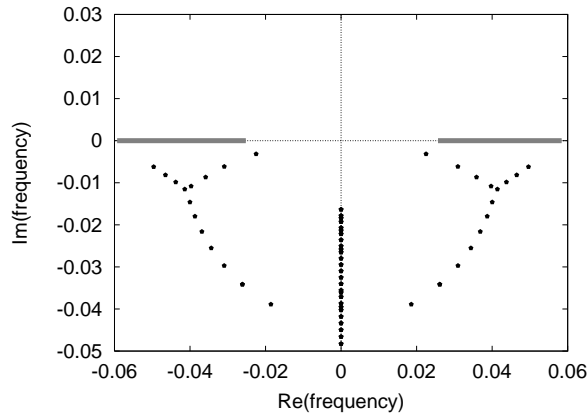}}
\caption{ A typical eigenvalue spectrum for a compressible 
($\gamma=\frac{5}{3}$)
resistive MHD case showing the complex frequencies of damped and growing
normal modes. In this low-pressure case, slow magnetosonic modes have
eigenfrequencies very close to the origin and are not shown. The plasma
model parameters are similar to the constant current case, but with a 
small shear: $m=1$, $\kappa=0.35$, $nq(r)=1.2\times(1-0.1r)$ and
$S=3 \times 10^{3}$. The ideal Alfv\'en continuum is represented by a 
grey line on the real axis.}
\label{figure:SimpleSpectrum}
\end{figure}

For general plasma configurations with shear, the resistive 
Alfv\'{e}n spectrum is
usually found to form a fork (e.g. figure \ref{figure:SimpleSpectrum}
or those in \cite{Davies} - \cite{KernerLerbinger})
The rather different shape of the spectral loci in
[figures \ref{figure:CompressTest2_gamma1}-(f)] is
a consequence of the equilibrium having an Alfv\'{e}n spectrum which
degenerates to a point.

\noindent The fork structure in the resistive MHD spectrum
has been qualitatively explained in terms of WKB
analysis by examining turning points within the plasma, see
\cite{DewarDavies} and  \cite{KernerLerbinger}. The fork
has three lines joining at a point below the ideal MHD continuum.
Two lines run between the intersection point and either end of the Alfv\'{e}n
continuum. The third line runs around approximately in a quarter circle to
touch the imaginary axis.
In a simple model with toroidal current density constant across the
plasma, there is an analytical solution
for the $\gamma \rightarrow \infty$ resistive MHD spectrum
\cite{StorerAnalytic}.
We show that a perturbed variant of this constant current model,
in which a slight
shear is given to the magnetic field, is still amenable to the
manipulations performed in \cite{StorerAnalytic}.
By introducing shear we produce a model
which has a finite width Alfv\'{e}n continuum, in which we might hope 
to recover the
generic fork structure found in compressible results.
We therefore solved this model
using WKB analysis to explain the qualitatively different spectrum.
In the remainder of this paper we set $\rho_{\perp}=\rho_{\parallel}=\rho$.

\section{WKB analysis of a small shear equilibrium in the limit
          $\gamma \rightarrow \infty$ }
\label{section:WKBAN}

The model case is derived from \cite{StorerAnalytic},
which considers a cylindrical plasma with a constant axial
field and no shear. This model has
been studied earlier in \cite{Tayler},\cite{Breus}.
The equations used for this analysis are those of linearised, resistive,
incompressible MHD, with $\gamma \rightarrow \infty$:

\begin{equation}
\rho \mu_0 \frac{\partial}{\partial t} ( \curl \vv ) =
    \curl (\Bv \dott \grad \bv + \bv \dott \grad \Bv)    ,
\label{eq:newtonslaw}
\end{equation}
and magnetic field given by Amp\'{e}re's law

\begin{equation}
  \frac{\partial \bv}{\partial t} =
    \nabla \cross (\vv \cross \Bv) - \curl ( \frac{\eta}{\mu_0} \curl
\bv ).
\label{eq:ampere}
\end{equation}

\noindent  The curl
of the equation of motion is taken in order to suppress the perturbed
pressure. Also, we specialise to an equilibrium state with no plasma
velocity.

\begin{figure}[htb]
\centerline{\epsfxsize=8cm \epsfbox{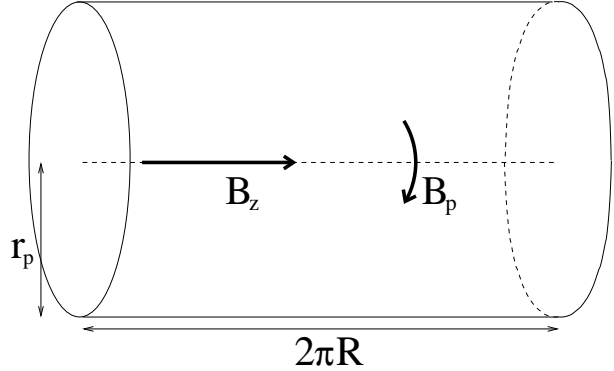}}
\caption{One period of the cylindrical model. }
\label{figure:Cylinder}
\end{figure}

\noindent The idea is to introduce the shear as a small quantity, of the same
order as the inverse wave number. The analysis is then the same 
as the shear-free case, up to two orders in the inverse wavenumber.
The radial dependence is included in the dispersion relation in the
radially varying quantities: $B_{z}(r)$, $\Bpol(r)$ and $q(r)$.
We first take the large wavenumber limit by ordering
$ \nabla \simeq O(1/\epsilon)$.
For significantly dissipative modes, maximal balance of
Amp\'{e}re's law (\ref{eq:ampere})
occurs for $\epsilon \simeq O(S^{-1/2})$. In a typical physical situation
we might have
$ S > 1000$ and thus $\epsilon < 0.03 $ is a good expansion
parameter.

\noindent The magnetic field is expressed as
$\Bv = \hat{z}B_z(r) + r \hat{\theta} \Bpol(r) $ with
$d ( \mbox{log}[\Bpol(r)])/d r$ and $ d (\mbox{log}[B_z(r)])/d r $ both
of $O(\epsilon) $, in order to satisfy the requirement of
small shear.
We again look at perturbations of the form
$\bv = \exp \left(
i m \theta - i \kappa z / r_p - i \omega
t \right) \bv\left(r \right) $
. For
convenience we set $\bv$ as $O(1)$ and this then implies
$\vv$ to be of $O(1)$ to complete the ordering.
By using the relations $\nabla \cdot \bv = \nabla \cdot \vv = 0 $,
equations (\ref{eq:newtonslaw})
and (\ref{eq:ampere}) can be reduced to:

\begin{equation}
\rho \mu_0 \frac{\omega}{\tau_A} ( \nabla \times \vv ) =
    \frac{B_p(r)}{r_p} [m - n q(r) ] \nabla \times \bv
     + \frac{2B_p(r)}{r_p^2}\kappa \bv + O(\epsilon)
\label{eq:fouriernewton}
\end{equation}
and
\begin{equation}
  -\frac{i \omega}{\tau_A} \bv = \frac{i B_p(r)}{r_p} [ m - n q(r) ] \vv
      - \frac{\eta}{\mu_0} (\nabla \times \nabla \times \bv ) + O(\epsilon).
\label{eq:fourierampere}
\end{equation}
\noindent
The safety factor $q(r)$ is given by
  ${r_p B_z(r)}/{R B_p(r)}$ and the non-dimensional resistivity $\eta=S^{-1}$.


\noindent In this form, the only differential operator is the curl
operator. This motivates us to look for solutions which are eigenfunctions
of this operator, suggesting the ansatz

\begin{eqnarray}
r_p \nabla \times \vv = \alpha \vv, \quad & r_p \nabla \times \bv = \alpha \bv,
\label{eq:curleq}
\end{eqnarray}

\noindent which solves equations (\ref{eq:fouriernewton}) and 
(\ref{eq:fourierampere})
provided
\begin{equation}
\alpha(r,\omega) = \frac{2[m-nq(r)]\kappa}
   {\frac{B_z(r)^2}{B_p(r)^2}i \omega [ i \omega
     -  S^{-1} \alpha(r,\omega)^2 ] +
                  [m -  n q(r)]^2}.
\label{eq:dispersionrelation}
\end{equation}

\noindent By taking the curl of equation (\ref{eq:curleq}) we get

\begin{equation}
\nabla \times \nabla \times \vv = -\nabla^2 \vv = \frac{\alpha^2}{r_p^2} \vv,
\label{equation:dispersionrelation}
\end{equation}
since the velocity is divergence-free. This implies a relation
for the $z$ component of $\vv$

\begin{equation}
\frac{1}{r}\frac{d}{dr} r \frac{d}{dr} v_z
      = -\left(\frac{\alpha^2}{r_p^2} + n^2 - \frac{m^2}{r^2} \right) v_z  .
\label{eq:waveeqnprec}
\end{equation}

\noindent This is amenable to standard WKB analysis if $\alpha$ is large,
and in this WKB limit equation (\ref{eq:waveeqnprec}) is equivalent to:

\begin{equation}
  \frac{d^2}{dr^2} \vv_z = -Q(r) \vv_z,
\label{eq:waveeq}
\end{equation}
\noindent with $Q(r)={\alpha^2}/{r_p^2} $.
This will break down near the origin ($r=0$)
where we will use a Bessel function matching.
Equation (\ref{eq:waveeq}) is solved approximately by:

\begin{equation}
  \vv_z \simeq a_{\mbox{out}}Q^{-\frac{1}{4}} e^{i\phi}
          + a_{\mbox{in}} Q^{-\frac{1}{4}} e^{-i\phi}
\end{equation}
where the amplitudes $a_{\mbox{out}}$ and $a_{\mbox{in}}$
are slowly varying functions, and
\begin{equation}
   \phi(r|c) =  \int_c^r Q^{\frac{1}{2}}(r') dr'
\end{equation}

\noindent Thus $\alpha / r_p $ is the radial wavenumber and equation 
(\ref{eq:dispersionrelation})
provides the dispersion relation.

\section{Characterising the Stokes points}

To find the WKB solutions, it is first necessary to examine the
structure of the dispersion relation in the plasma region.
In particular, singularities and zeros and
the associated branch structure of the dispersion relation
must be examined. Branch points of the dispersion relation are
known as {\it Stokes points}.
The dispersion relation (\ref{eq:dispersionrelation})
can be written as a cubic equation in $\alpha$, with the coefficients
as functions of $q(r)= {r_p B_z(r)}/{R B_p(r)}$, i.e.

\begin{equation}
\alpha^3 \frac{1}{S}{\left(\frac{q(r) R}{r_p} \right)}^2
+\alpha{\left( [m - n q(r)]^2
               -{\left(\frac{q R}{r_p} \right)}^2 \omega^2 \right)}
  = 2 [m - n q(r)] \kappa.
\label{equation:cubiccomplicated}
\end{equation}

\begin{figure}[htb]
\centerline{\epsfxsize=8cm \epsfbox{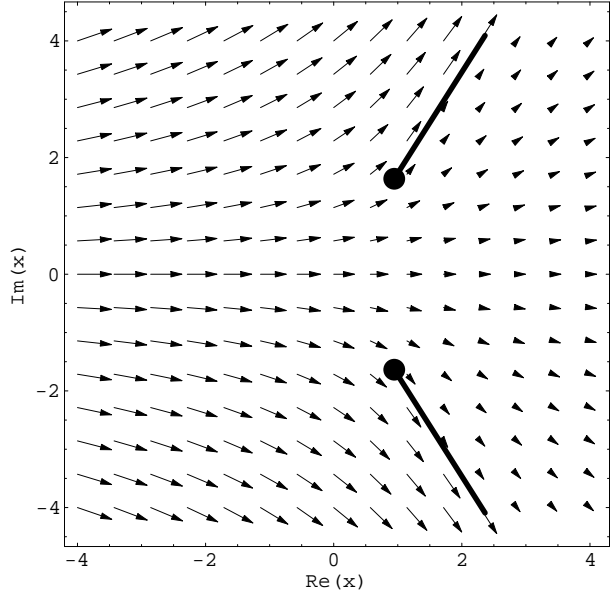}}
\caption{One branch of the multivalued function $\bar{\alpha}(x)$
($\xi = -1$)
shown on the complex plane as a Polya plot. Branch cuts
are indicated as thick lines. The vector
($\mbox{Re}[\bar{\alpha}],
   -\mbox{Im}[\bar{\alpha}]) $ is displayed on
a grid.}
\label{figure:SingleBranch}
\end{figure}

\noindent We would like to discover the singularity structure of our dispersion
relation. Solving equation 
(\ref{equation:cubiccomplicated}) for $\alpha$ leads to very ungainly
equations and proves not to be
enlightening, so
we look for a simpler relation which will be
topologically equivalent.
Let us consider the case where there is no magnetic surface resonant
with the perturbation. In this case we have $ [ n q(r) - m ] \neq 0$
within the plasma, and assuming also $ q(r) \neq 0 $
then we can divide through the equation by $2[ m-n q(r)]\kappa $
and introduce a new variable $\bar{\alpha}$ so that

\begin{equation}
{\bar{\alpha}}^3+x(r) \bar{\alpha} = 1,
\label{eq:simplecubic}
\end{equation}
with
\begin{equation}
\bar{\alpha} \equiv \alpha \frac{
         \{2 [n q(r) - m] \kappa S\}^{\frac{1}{3}} }
         {
          {\left({q(r) R}/{r_p} \right)}^{\frac{2}{3}}
         }
\end{equation}
and
\begin{equation}
  x(r) \equiv  \frac{ \left\{
            [n q(r) - m]^2 - \left[ {q R}/{r_p} \right]^{2} \omega^2
        \right\} } 
         { { \left\{
          2 [n q(r) - m] \kappa S {\left( {q R}/{r_p} \right) }
        \right\} }^\frac{2}{3} }.
\end{equation}
The solution of equation (\ref{eq:simplecubic}) for $\bar{\alpha}$ is

\begin{equation}
\bar{\alpha} = - \xi \frac{2^{\frac{1}{3}}\,x}
      {{\left( 27 + {\sqrt{729 + 108\,x^3}} \right) }^{\frac{1}{3}}}
       + \xi^{*} \frac{{\left( 27 + {\sqrt{729 + 108\,x^3}} \right) }^
      {\frac{1}{3}}}{3\,2^{\frac{1}{3}}},
\label{eq:cubicsolution}
\end{equation}
where $\xi$ is one of the cube roots of $-1$:

\begin{eqnarray}
    \xi = -1 ,
    & \frac{1 + i \,{\sqrt{3}}}{2},
    & \frac{1 - i \,{\sqrt{3}}}{2} .
\end{eqnarray}
We consider $x$ as a new radial variable.
The function $\bar{\alpha}$ is represented graphically
by the Polya plot in figure
\ref{figure:SingleBranch}.

\noindent By inspection of the form of equation (\ref{eq:cubicsolution})
we have candidates for branch points at the three roots of $x^3=-27/4$.
However, not all of these candidate branch
points are true branch points, as suggested by figure 
\ref{figure:SingleBranch}.
This can be seen in the $\xi = -1 $ case, where
we have, from equation (\ref{eq:simplecubic})

\begin{equation}
x(\bar{\alpha}) = \frac{1-{\bar{\alpha}}^3}{\bar{\alpha}},
\label{eq:bequation}
\end{equation}
which can be considered as a local inverse of equation (\ref{eq:cubicsolution}).

\noindent Let us consider the neighbourhood of
  $x = -\sqrt[3]{{27}/{4}} $
(using the principal value, so $x$ is a negative real). We might expect
a branch point here from the structure of equation (\ref{eq:cubicsolution}).
At $x = -\sqrt[3]{27/4} $ we have
$\bar{\alpha} = 2^{2/3}$.
However, there is a neighbourhood around $\bar{\alpha} = 2^{2/3}$
where equation 
(\ref{eq:bequation}) is analytic and has a non-zero derivative, and
therefore the function has an analytic inverse around
$x = -\sqrt[3]{27/4} $.
There obviously cannot be a branch point in an analytic region.
The other two candidate points are true branch points.
It similarly follows that each of the other cases of equation 
(\ref{eq:cubicsolution}) have only two branch points each.
Note that around a Stokes point at some position $x_0$,
we do not have $\alpha \propto \sqrt{x - x_0}$,
as is typical for many WKB analyses \cite{BerkPfirsch}.
Instead, we have $\alpha \simeq C + D \sqrt{x - x_0} $.

\section{Phase matching: a solution near the singularities}

In order in proceed with WKB analysis, we need to determine the behaviour of
solutions near the Stokes points, the branch points
of the dispersion
relation. In the neighbourhood of the branch point, we approximate the
dispersion relation by:

\begin{equation}
  Q(x) \simeq 1 + A x^{\frac{1}{2}} .
\end{equation}
This is unlike the more usual situation in
WKB analysis where $Q(x) \simeq x $
around the Stokes points.
The simplest treatment of the phase matching follows from considering
  $ A \ll 1 $ in which case the $ A = 0 $ case can be used as a zeroth
order solution in a region around the Stokes point. Note that for $A=0$,
the dispersion relation is independent of $x$ and there is no reflection
of the wave. As we will see, as $ A \rightarrow 0 $ the reflectivity goes
to zero. The transmitted part of the wave will be decaying for finite $A$,
so that we have partial absorption of the travelling wave.

\noindent Our wave equation is
\begin{equation}
\frac{d^2 y}{d x^2} = -(1 + A x^{\frac{1}{2}} ) y,
\label{eq:compwaveeq}
\end{equation}
with an $A=0$ solution
\begin{equation}
y_0 = e^{\pm i x},
\end{equation}
which motivates the substitution
\begin{equation}
y = e^{-i x + u(x) }.
\label{eq:expsub}
\end{equation}
The other choice of sign leads to a
second solution to the equation, which is
growing for $x \rightarrow - \infty $.
Substitution of equation (\ref{eq:expsub}) into equation (\ref{eq:compwaveeq}) leads to
\begin{equation}
A \sqrt{x} - 2 i \frac{d u}{d x} + \left( \frac{d u}{d x} \right)^2
             + \frac{d^2 u}{d x^2}=0.
\end{equation}

\noindent We are looking for small departures from 
the $A=0$ solutions and in
this case we can choose ${d u}/{d x} \ll 1 $ so that to first order

\begin{equation}
A \sqrt{x} - 2 i \frac{d u}{d x} + \frac{d^2 u}{d x^2}=0,
\end{equation}
from which we can find $u'(x)$


\begin{equation}
  {{u'(x)} =
      {e^{2\,i \,x}\,C -
        \frac{i }{8}\,A\,
         \left[ 4\,{\sqrt{x}} +
           \,e^{(2 \,x - \pi/4) i}\,
            {\sqrt{2\,\pi }}\,
            \mbox{erf}\left(
            \,e^{- 3 \pi i/4}   {\sqrt{2 x}}\right) \right] }}.
\label{eq:uprimeprecursor}
\end{equation}

The coefficient of integration, $C$, must now be chosen
  such that we can match the
solution on the left-hand side of the origin to the evanescent WKB
solution.
We have required $u'(x) \ll 1 $, so an oscillatory $u(x)$ can be
modelled as $ \epsilon e^{2 i x} $ with $\epsilon \ll 1 $
(plus a constant which can be safely
ignored) in which case:

\begin{equation}
y(x) = e^{- i x  + u(x) } = e^{-i x}e^{\epsilon e^{2 i x }}
  \simeq e^{-i x }\left( 1 + \epsilon e^{2 i x } \right) =
e^{-i  x } + \epsilon e^{i x}.
\end{equation}
\noindent These correspond to the WKB solutions, which
  are approximately of the form $c_1 e^{-i x}+ c_2 e^{i x}$
near the origin. We require that the WKB solution matched
on the left-hand side have $c_2 = 0 $
because the corresponding term grows exponentially for large negative
$x$.
We therefore must have $\epsilon \rightarrow 0 $ as $x \rightarrow - \infty $.
Using the asymptotic expansion of
$\mbox{erfc}=1-\mbox{erf}$, as
given by equation 7.1.23 of \cite{AbrStegun}, 
we find the $x \rightarrow - \infty $ limit of equation (\ref{eq:uprimeprecursor}),
allowing us to express this matching condition as:
\begin{equation}
  C =  \frac{A}{8} (1+i) \sqrt{\pi} .
\end{equation}
Then
we have a solution for $y$ which is asymptotically of the form:

\begin{equation}
y = P(x) \left( e^{ -i x} + e^{ i x }
   \frac{A \sqrt{\pi}(1-i)}{4}
  \right)
\end{equation}
for $x \gg 1 $, with $P(x)$ a slowly varying function.
The phase matching condition is given by finding the nodes of these
waves, which fixes the WKB phase at $ x=0 $:

\begin{equation}
\phi_0 = \frac{- i }{4}\,
    \log(\frac{(1+i)A\sqrt{\pi}}{8}).
\end{equation}

\section{ Finding wavemodes }

Global modes are found in the usual way:
we look for paths $C$ in the complex plane joining the
axis and boundary where
$\int_{B} \alpha (x) dx $
  is real for any sub-path $B$ of $C$. These paths will
be WKB solutions if the integral $\int_{C} \alpha(x) dx=
\int_{[0,1]} \alpha(x) dx $ which can be guaranteed if there are no
singularities of our differential equation coefficients in the region.
In particular, this requires that the circular path $C-[0,1]$ does not
enclose any Stokes points.
The quantisation condition is supplied by requiring the correct
behaviour at boundaries. At the origin the WKB wavemode must be matched
to a Bessel function, and this gives the condition
  $\phi |_{x=0} = \pi (1/4 + m/2) $. At the outer boundary of the plasma,
we require $ v_r = 0 $ (fixed plasma boundary),
which leads to $\phi |_{x=1} = \pi/2 $.

\noindent Localised modes proceed from the axis or outer boundary of the
plasma and propagate along ray trajectories (which will in this case be
anti-Stokes lines) to a Stokes point. They are then evanescent past this
point, so it must be possible to draw a path connecting the relevant
Stokes point to the other boundary
without crossing a Stokes line.
For Stokes points of the form $Q(x) = a + b x^{\frac{1}{2}} $, which
are present in this analysis, we have a complex phase matching criterion.
The phase integral between the Stokes point and the boundary is then
required
to be complex for matching to occur.
This means that we cannot follow anti-Stokes lines,
along which the phase is real,
exactly to join the boundary and the Stokes point.
The complex portion of the phase leads to a correction to the path,
which must be taken into account.

\section{Application of the WKB method to the small shear incompressible
case}

For explicit studies, we use a small shear test case:

\begin{eqnarray}
B_{p}(r)=\frac{B_{p0}}{1- r \delta}=\frac{\frac{7}{24}}{1- 0.1 r}, \\
B_{z}(r)=1,
R=\frac{20}{7},
n=10,
m=1 ,
r_{p}=1,
 q(r) = 0.12(1-0.1r)
\label{eq:numerwkb}
\end{eqnarray}
The WKB trajectories in the complex $r$ plane were
determined numerically, and
several loci found by finding appropriate paths in the plane,
as in \cite{DewarDavies}.
The loci can be characterised by
the branch of the dispersion relation which they lie on, and whether
the corresponding wave modes are fully global modes or have a turning
point inside the plasma.

\noindent Although there are three branches of the dispersion relation,
on one branch it is never
possible in practice to form global modes: the
rays inevitably escape towards complex infinity. The
other two branches then produce
the two forks.

\begin{figure}[htb]
\centerline{\epsfxsize=8cm \epsfbox{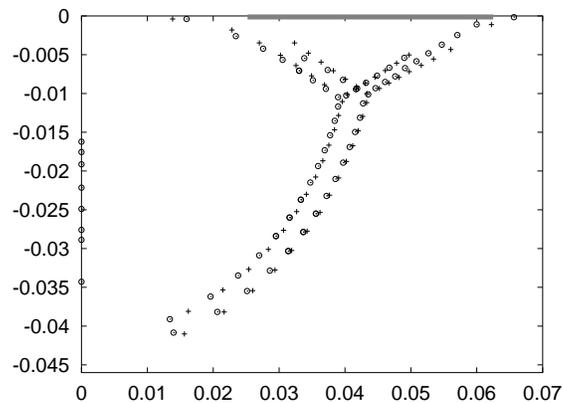}}
\caption{The resistive spectrum from numerical solution of the MHD 
equations (circles) compared to
  the WKB result (plus signs). The ideal Alfv\'en continuum is
represented by a grey bar on the real axis.
Plasma parameters are given by equation (\ref{eq:numerwkb}) and
$S=3 \times 10^{4}$.}
\label{figure:SmallShearNumericalSpectrum}
\end{figure}

\noindent The eigenvalues are displayed in figure
\ref{figure:SmallShearNumericalSpectrum}, together with the numerical result
from a code based on \cite{Gruber}. The
spectrum is qualitatively similar to a fork structure, but also shares
the features of the original simple model. Note that the double loci
(running parallel to each other in an arc)
are still present in this model.

\noindent The nature of the difference between the two branches of 
the double locus
can be seen in equation (\ref{eq:curleq}), and the form of the dispersion
function for large enough $\alpha$. Here we have two solutions for 
$\alpha(\omega)$
such that $\alpha_1 \simeq -\alpha_0$, and the two WKB solutions consist
of waves of opposite helicity. Finite pressure gradients in this
equilibrium result in waves of opposite helicity having slightly
different frequencies.

\section{Effects of the $\gamma \rightarrow \infty$ approximation}

The reason why we see a pair of loci in \ref{figure:CompressTest2_gamma1000},
rather than the single locus usually depicted for
compressible spectra (e.g. figure \ref{figure:CompressTest2_gamma1})
is that in this incompressible model (the limit $\gamma \rightarrow \infty$)
there are two
classes of wavemodes present which
can be excited at the Alfv\'{e}n frequency. In a uniform field
these wavemodes are degenerate: they oscillate at the same frequency.
However the two frequencies are split when the plasma contains currents
perpendicular to the magnetic field
(i.e. in non-force-free plasmas).
In the compressible model at low $\beta$, these two degrees of freedom
correspond to the slow (magnetosonic) and Alfv\'{e}n wavemodes and the
ratio between slow frequencies and Alfv\'{e}n frequencies is of
order $\beta^{1/2}$.

\noindent Force-free models are important special cases, in which pairs of
loci of eigenvalues coincide.
The $\gamma \rightarrow \infty$ approximation will still
result in unphysical eigenmodes.
We note the paper of Ryu and Grimm \cite{RyuGrimm}, which uses
this incompressibility assumption
to analyse a case with finite pressure gradients where
we should see a double locus structure. We nevertheless see a
simple fork structure.
We believe that the splitting effect is rather
small in this case, so that what looks like one fuzzy locus
is in fact a double locus.

\section{Conclusions}

In plasma physics an assumption of incompressibility
is often justified because the parallel dynamics of the plasma
and the fluid compression across the field are much less 
important than the forces due to the magnetic field. For example,
incompressibility does not generally affect ideal MHD marginal
stability (but this does not extend to resistive MHD \cite{CompResBall}).

\noindent Two incompressible resistive MHD models 
were compared with the physical
compressible model by analysis of their spectra. For the first model, where 
the ratio of specific heat is taken to infinity, we expect 
from local analysis to find two
types of wavemodes present at the Alfv\'{e}n frequency.
In the second model where we again set $\gamma \rightarrow \infty$, 
the parallel
plasma inertia is set to zero, and we expect only one Alfv\'{e}nic mode to be
present in the spectrum, corresponding to the physical case. Numerical
computation of the spectra of a magnetic shear-free plasma 
confirms that the second model reproduces
most of the eigenmodes associated with the Alfv\'{e}nic model correctly.
The first model has twice as many modes present at the Alfv\'{e}n
timescale.

\noindent It is noted that in general
most of the modes resolved do not correspond to Alfv\'en modes
and have no physical significance.
The shape of an incompressible spectrum for a more general model,
with shear present, was determined numerically and by WKB analysis.
The unusual nature of the local dispersion
relation leads to a complex structure of loci. 
The resulting spectrum included the `double locus' of the zero shear
model and also demonstrated the fork structure that is seen generically
for stable resistive MHD spectra. 

\noindent There are many qualitative features of the resistive Alfv\'en
spectrum that can be reproduced by simply setting
$\gamma \rightarrow \infty$.
Unfortunately, physical wavemodes and
frequencies are not well modelled in this approximation.
The stable part of the ideal Alfv\'en
spectrum is irreparably mixed with spurious modes in this limit.
However, by using an anisotropic mass density tensor, an incompressibility
constraint can be introduced while preserving the Alfv\'en modes.

\vspace{1cm}

\noindent {\bf{Acknowledgments}}

\noindent This work 
has been supported by the Flinders Institute for Science and 
Technology, the Australian Institute for Nuclear Science and 
Engineering and the Australian Partnership for Advanced 
Computing.

\vspace{1cm}

\noindent {\bf{References}}

\

\bibliographystyle{aipprocl}

\bibliography{resistive}

\end{document}